\documentclass{article}
\usepackage{spconf,amsmath,graphicx, url, enumitem, setspace}
\usepackage{makecell}
\usepackage{amssymb}
\usepackage{multirow}
\usepackage{booktabs}
\usepackage[dvipsnames]{xcolor}

\title{The Potential of Neural Speech Synthesis-Based Data Augmentation\\for Personalized Speech Enhancement}
\name{Anastasia Kuznetsova, Aswin Sivaraman, Minje Kim\thanks{This material is based upon work supported by the National Science Foundation under Grant No. 2046963.}}
\address{Indiana University, Luddy School of Informatics, Computing, and Engineering, Bloomington, IN, USA}
\begin{document}
\ninept
\maketitle
\begin{abstract}
With the advances in deep learning, speech enhancement systems benefited from large neural network architectures and achieved state-of-the-art quality. However, speaker-agnostic methods are not always desirable, both in terms of quality and their complexity, when they are to be used in a resource-constrained environment. One promising way is personalized speech enhancement (PSE), which is a smaller and easier speech enhancement problem for small models to solve, because it focuses on a particular test-time user. To achieve the personalization goal, while dealing with the typical lack of personal data, we investigate the effect of data augmentation based on neural speech synthesis (NSS). In the proposed method, we show that the quality of the NSS system's synthetic data matters, and if they are good enough the augmented dataset can be used to improve the PSE system that outperforms the speaker-agnostic baseline. The proposed PSE systems show significant complexity reduction while preserving the enhancement quality.
\end{abstract}
\begin{keywords}
personalized speech enhancement, data augmentation, speech synthesis
\end{keywords}
\section{Introduction}
\label{sec:intro}

Designing algorithms for speech enhancement (SE) is a long-standing research problem 
in which the current state-of-the-art methods use deep neural networks (DNNs) \cite{XuY2014ieeespl,, PascualS2017segan, WangDL2018ieeeacmaslp, HuY2020dccrn, EskimezSE2021personalized}.
DNN-based noise suppression algorithms typically utilize a training set prepared by artificially mixing arbitrary noise sounds with clean speech signals from many different speakers.
As a result, fully-trained SE systems attempt to enhance any speech within a given input mixture.
These models can generalize to the unseen test speakers if the model's computational capacity is large enough to encompass the variations found in thousands of speakers and noise types.
Therefore, generalist models come at the cost of increased test-time complexity.

Recent studies have explored methods for developing target speaker extraction models, which employ a conditioning mechanism to focus on the target speaker out of a mixture of multiple talkers and noise sources. 
A straightforward strategy for targeting on a particular speaker is to condition the model with a speaker embedding \cite{GiriR2021interspeech}, inferred from around 5 to 10 seconds of enrollment data (i.e., clean speech) of the target speaker \cite{ZhouT2021resnext}.
Likewise, embedding-based conditioning frameworks effectively merge the tasks of speaker extraction with speech enhancement \cite{WangQ2018voicefilter, EskimezSE2021personalized}.

A certain type of personalized speech enhancement (PSE) method seeks the potential benefits of personalizing the model beyond just better performance. Its goal is to reduce the model complexity while preserving the enhancement quality. It is possible because 
on the one hand, learning to enhance a single speaker is simpler than enhancing many speakers, allowing for less complex models.
Therefore, PSE may be seen as a model compression mechanism, as more-efficient PSE models may be deployed in place of larger speaker-agnostic SE models without sacrificing performance \cite{KimSW2021waspaa, SivaramanA2020interspeech, SivaramanA2021waspaa, SivaramanA2022jstsp}.
On the other hand, personalization is a challenging optimization task unless it gets help from the enrollment procedure that can provide the target speaker's information. However, the reference signals acquired in this way come at the cost of untrustworthy recording quality and privacy concerns.
Consequently, reducing the amount of the target speaker's clean speech is an additional constraint for PSE systems toward data efficiency.

In this work, we investigate a novel data augmentation strategy for PSE systems using neural speech synthesis (NSS) data.
We consider the scenario in which a PSE model has access to a very limited amount of speech data from the target speaker, e.g., 5 seconds of clean reference speech. We investigate whether this amount is good enough for some off-the-shelf NSS systems to generate clean speech signals, while preserving the speaker's identity. In this constrained condition, their synthesis quality must vary depending on the model's performance, and there are various ways to evaluate the perceptual quality of synthesized speech, such as naturalness, intelligibility, personality, emotions, etc. 
Hence, our goal is to analyze the correlation between the varying quality of synthesized speech and the PSE performance using this augmented dataset. 
We compare two different NSS-augmented datasets and their usability on PSE models. Furthermore, we also compare the proposed approach against non-personalized SE models as well as self-supervised PSE models  \cite{SivaramanA2022jstsp}.
Our results show that NSS dataset augmentation is useful for PSE, especially in cases where the model is too small to generalize well to the test speaker.
We also observe that the quality of synthesized speech impacts PSE performance.
Our findings are consistent across different model sizes, where NSS-augmented PSE models outperform speaker-agnostic models.
This suggests that sub-optimal NSS synthesis is still advantageous in the context of personalizing speech enhancement systems with a simpler training framework compared to existing methods.

\begin{table*}[t]
\centering
\resizebox{.8\textwidth}{!}{%

\begin{tabular}{cccclc}

\toprule
\textbf{Set} & \textbf{Subset} & \textbf{Duration} & \textbf{Quantity} & \textbf{Description}   & \textbf{Corpus}\\
\midrule
\multirow{2}{*}{$\mathbb{G}$} & $\mathbb{G}_{\text{tr}}$ & 443 h            & 1132 spkrs   & \multirow{2}{*}{Clean speech from many anonymous speakers} & \multirow{2}{*}{LibriSpeech \cite{PanayotovV2015Librispeech}} \\
& $\mathbb{G}_{\text{vl}}$  & 8h    & 20 spkrs &    &   \\
\midrule
\multirow{2}{*}{$\mathbb{S}^{(1:20)}$}        & $\mathbb{S}^{(1:20)}_{\text{tr+vl}}$ & $5\times 30$ sec/spkr & 20 spkrs          & Target speakers set used for synthesis  & \multirow{2}{*}{LibriSpeech \cite{PanayotovV2015Librispeech}}  \\
    & $\mathbb{S}^{(1:20)}_{\text{te}}$                    & 30 sec/spk               & 20 spkrs          & Clean speech target speaker set only used for evaluation  &  \\
    \midrule
\multirow{2}{*}{$\mathbb{S}^{(21:26)}$}        & $\mathbb{S}^{(21:26)}_{\text{tr+vl}}$                  & 3 sec/spkr & 6 spkrs   & Target speaker set used for synthesis                             &              \multirow{2}{*}{LibriSpeech \cite{PanayotovV2015Librispeech}}   \\
    & $\mathbb{S}^{(21:26)}_{\text{te}}$  & $\sim$8 sec/spk          & 6 spkrs           & Clean speech target speaker set only used for evaluation   &  \\
\midrule
\multirow{2}{*}{$\Tilde{\mathbb{S}}^{(1:20)}$}  & $\Tilde{\mathbb{S}}^{(1:20)}_{\text{tr}}$ & 60 sec/spkr              & 20 spkrs          & \multirow{2}{*}{Synthesized target speaker utterances (YourTTS)}  &    \\
 & $\Tilde{\mathbb{S}}^{(1:20)}_{\text{vl}}$   & 30 sec/spkr   & 20 spkrs          &   &   \\
 \midrule
\multirow{2}{*}{$\Tilde{\mathbb{S}}^{(21:26)}$} & $\Tilde{\mathbb{S}}^{(21:26)}_{\text{tr}}$                 & 21 sec/spkr              & 6 spkrs           & \multirow{2}{*}{\begin{tabular}[c]{@{}l@{}}Synthesized target speaker utterances\\(YourTTS, AudioLM)\end{tabular}}  & \\
& $\Tilde{\mathbb{S}}^{(21:26)}_{\text{vl}}$                  & 7 sec/spkr               & 6 spkrs   &       &  \\
\midrule
\multirow{3}{*}{$\mathbb{N}$}         & $\mathbb{N}_{\text{tr}}$  & 5h                       & 616 noises        & \multirow{2}{*}{Injection noises used during SE and PSE training}   & \multirow{3}{*}{MUSAN \cite{SnyderD2015MUSAN}}       \\
& $\mathbb{N}_{\text{vl}}$ & 0.5h & 60 noises         &   &    \\
                           & $\mathbb{N}_{\text{te}}$ & 0.5h & 60 noises         &
\makecell[l]{Injection noises not seen during training, \\ used to prepare speaker-specific test sets}                           &         \\
\midrule
$\mathbb{T}$ &  & & 88156 sentences & \makecell[l]{Sentences used for synthesis in both \{tr, vl\}  \\ partitions of $\Tilde{\mathbb{S}}^{(1:20)}$ and $\Tilde{\mathbb{S}}^{(21:26)}$.}

& VCTK \cite{Yamagishi2019CSTRVC}\\
\bottomrule
\end{tabular}
}
\caption{Description of the datasets used in experiments.}
\label{tab:data}
\end{table*}

\section{Proposed Method}
\label{sec:proposed_method}

\textbf{The generalists}: A fully-supervised framework for training SE models defines a large set $\mathbb{G}$ of clean utterances from many anonymous speakers. They are mixed with various noise signals sampled from noise corpus $\mathbb{N}$ at random signal-to-noise ratios (SNRs) to simulate arbitrary contaminated speech, i.e., $x = s + n$ where $s \in \mathbb{G}$ and $n \in \mathbb{N}$. The SE model is a mapping function $f(\cdot)$ with trainable parameters $\mathcal{W}_{\text{SE}}$ that aims to recover $s$ from $x$, i.e., $f(x;\mathcal{W}_{\text{SE}}) = y \approx s$. 
Our experiments use negative signal-to-distortion ratio (SDR) \cite{VincentE2006ieeeaslp}  as the loss function for the SE system: 
\begin{equation}
    \label{eq:sdr}
    \mathcal{L}_{\text{Neg-SDR}} (\hat{v} \| v) = -10 \log_{10} \left[ \frac{\sum_t (v_t)^2}{\sum_t(v_t - \hat{v}_t)^2} \right ],
\end{equation}
where $v$ is the clean signal and $\hat{v}$ is the estimated signal. We refer to such models as generalists and use them as a baseline. 

\textbf{PSE using NSS}: The \textit{transfer learning} approach can introduce a bias towards a particular speaker. As opposed to random initialization, transfer learning deems to be highly beneficial for specialists if the speaker-specific clean speech is available for finetuning, which we denote by $\mathbb{S}^{(i)}$ with a speaker index $i$. We assume $|\mathbb{S}^{(i)}| \ll |\mathbb{G}|$ due to any technical challenges in acquiring clean recordings or privacy concerns of the user, e.g., a few seconds. Therefore, we propose the data augmentation technique for PSE using neural speech synthesis (NSS) systems. Here, we assume that the NSS system is a text-to-speech (TTS) system that can generate any utterances that sound similar to the target speaker.

\textbf{Data augmentation via NSS}: Our goal in using the NSS systems is to generate as many new utterances per speaker as needed using an NSS model $g(\cdot)$ with pretrained parameters $\mathcal{W}_{\text{NSS}}$. A sentence is sampled from a large set $t \in \mathbb{T}$ fed to the NSS model as $\Tilde{s} = g(t; \mathcal{W}_{\text{NSS}} | s)$, where $s$ works as a condition to preserve target speaker's personality as $s  \in \mathbb{S}^{(i)}$. Thus, we get a synthesized set of target speaker utterances $\Tilde{\mathbb{S}}^{(i)}$ of any predefined size. In training phase $\Tilde{s}$ is used the same way as $s$ to construct a noisy mixture $\Tilde{x} = \Tilde{s} + n$. We define the task of data augmented PSE as $f(\Tilde{x};\mathcal{W}_{\text{PSE}}) = \Tilde{y} \approx \Tilde{s}$. 
We hypothesize that the quality of the speech synthesis system affects the performance of the PSE, where two main factors are involved in defining the NSS systems' performance: the general quality of the speech signal and personality. While these two concepts are not straightforward to quantify, we empirically show that the two NSS systems, in comparison, are with different performances, and they are correlated to their usefulness in the PSE task. 
Our experiments address the following research questions: (a) does the quality of an NSS system impact its usefulness towards PSE? (b) for each of the tested NSS systems, how much generated data is needed in order for a transfer learning-based PSE system to perform comparably to large generalist models? (c) how much lossless compression can we achieve with NSS-augmented PSE models in comparison with generalist SE models?

\section{Experimental Setup}
In our experiments, we take the baseline speaker-agnostic generalist model proposed in \cite{SivaramanA2022jstsp} and finetune it with utterances synthesized by two off-the-shelf NSS models that exhibit varying performances. We repeat the finetuning process for multiple target speakers to comprehensively assess the proposed method. The speech enhancement  model complexity between four preset sizes in order to investigate the merit of PSE in terms of model compression. 

\subsection{NSS Models}

Our first NSS system is YourTTS \cite{casanova2022yourtts}, a multi-lingual multi-speaker TTS model composed of a transformer-based encoder, a normalizing flow decoder, and a HiFi-GAN vocoder. Second, we choose AudioLM \cite{borsos2022audiolm}, which is a novel auto-regressive speech synthesis system based on a language modeling (LM) approach. As opposed to YourTTS, AudioLM does not require textual data to synthesize speech, because the LM generates a word sequence on the fly. Instead of attempting to reproduce this model, whose pretrained version is unavailable, we use the small number of examples published in the authors' website\footnote{\url{https://google-research.github.io/seanet/audiolm/examples/}}. Although these examples are a small dataset, their relative higher quality provides an interesting comparison point to the large quantity of YourTTS results.

\subsection{Datasets}
\label{sub:data}

Table \ref{tab:data} describes all the datasets used in our experiments. The subscripts `tr', `vl', and `te' denote training, validation, and test subsets, respectively. We chose to work with two different subsets of speakers $\mathbb{S}^{(1:20)}$ and $\mathbb{S}^{(21:26)}$ for better comparison. $\mathbb{S}^{(1:20)}$ contains 20 speakers from LibriSpeech's \cite{PanayotovV2015Librispeech} \textit{train-clean-100} subset 
to match the training setup in \cite{SivaramanA2022jstsp}. As for the second set $\mathbb{S}^{(21:26)}$, we are based on the high-quality audio samples generated by \mbox{AudioLM} available online. Since the identity of those speakers is not fully known, we work with only 6 identifiable speakers from \textit{test-clean}, $\mathbb{S}^{(21:26)}$. 

We generate the augmented version $\Tilde{\mathbb{S}}^{(1:20)}_\text{YourTTS}$ using YourTTS. For a given speaker ID $i$, the system takes a 5-second-long clean reference audio of the target speaker from $s \sim\mathbb{S}^{(i)}$ and a random sentence $t\sim \mathbb{T}$, which are the input pair $(s, t)$ to YourTTS that generates $\Tilde{s}$. We repeat the process until we collect a pre-defined duration of speech for the target speaker. 

Data augmentation for the second subset ${\mathbb{S}}^{(21:26)}$ is conducted using both NSS systems. First, as for AudioLM, we conveniently repurpose their publicly available synthesis results. The 7-second-long AudioLM utterances in $\Tilde{\mathbb{S}}_\text{AudioLM}^{(21:26)}$ are generated from a 3 seconds of audio prompt; four such synthesized examples are available per speaker, making 28 seconds of synthesized audio in total.
We generate another augmented set $\Tilde{\mathbb{S}}_\text{YourTTS}^{(21:26)}$ using the same 3-second long prompt, but this time we can synthesize as long utterances as we want because we have access to the pretrained model. All the utterances are resampled to 16 kHz.

For noise subsets we use \textit{sound-bible} partition of MUSAN~\cite{SnyderD2015MUSAN} only for test-time mixtures $\mathbb{N}_{\text{te}}$, 60 noise files from  \textit{free-sound} folder are set aside for validation mixtures $\mathbb{N}_{\text{vl}}$ and the rest of the signals from \textit{free-sound} are used for training time noisy mixtures $\mathbb{N}_{\text{tr}}$. Mixturre SNR is chosen from $[-5, 5]$ dB at random.

\subsection{Implementation}
\label{sub:implementation}
All models in our experiment are based on the well-known monoaural time-domain source separation DNN, ConvTasNet (CTN) \cite{LuoY2019conv-tasnet}. Following \cite{SivaramanA2022jstsp}, we define four different sizes of the model: large, medium, small and tiny. With each size variant the number of channels in the bottleneck module and convolutional blocks is reduced by factor of 2. Consequently, the number of trainable parameters in each model is 138.8K, 224.1K, 437.8K, and 1M, from the tiny to the large models, respectively. We assume smaller models are more suitable for on-device speech enhancement.



The generalist models are pretrained using the Asteroid implementation of ConvTasNet \cite{ParienteM2020asteroid} as in \cite{SivaramanA2022jstsp} and then finetuned using the proposed augmented datasets. We use Adam optimizer \cite{KingmaD2015adam} with a learning rate of $1\mathrm{e}{-5}$ and batch size of 8. After seeing 500 mixtures the model is validated on a fixed set of mixtures depending on the size of the training set: for  $\Tilde{\mathbb{S}}^{(1:20)}$ the validation set consists of 100 mixtures, for $\Tilde{\mathbb{S}}_\text{AudioLM}^{(21:26)}$ and $\Tilde{\mathbb{S}}_\text{YourTTS}^{(21:26)}$ the validation set consisted of 30 mixtures. 
The training continues until there is no validation SDR improvement for 5000 mixtures.
To show the improvement of finetuned models we report SDR, PESQ \cite{RixA2001pesq} and eSTOI \cite{TaalC2010icassp} for both test and validation sets.

\begin{table}[t]
\centering
\centering
\resizebox{.75\columnwidth}{!}{%
\begin{tabular}{@{}lrr@{}}
\toprule
\multicolumn{1}{c}{\textbf{Subset}}  & \textbf{MOS (est.)} & \textbf{Cosine Similarity} \\ \midrule
$\mathbb{S}^{(1:20)}_\text{YourTTS}$; 60 sec. &     3.78    & 0.80       \\
\midrule
$\mathbb{S}^{(1:20)}_\text{YourTTS}$; 120 sec. &    3.81   & 0.80           \\
\midrule
$\mathbb{S}^{(21:26)}_\text{AudioLM}$; 21 sec. &     \textbf{4.35} & \textbf{0.96}             \\
\midrule
$\mathbb{S}^{(21:26)}_\text{YourTTS}$; 21 sec. &    4.18     & 0.87       \\
\midrule
$\mathbb{S}^{(21:26)}_\text{YourTTS}$; 60 sec. &     4.12      & 0.88         \\ \bottomrule
\end{tabular}
}
\caption{Quality evaluation of synthesized speech for YourTTS and AudioLM generated sets.}
\label{tab:synth_quality}
\end{table}

\begin{table*}[t]
\centering
\resizebox{.81\textwidth}{!}{%
\begin{tabular}{@{}c|c|rrrr|rrrr@{}}
\toprule
\textbf{Experiment \#1}                 &{\textbf{Size}} & \multicolumn{1}{c}{\textbf{SDRI (te)}} & \multicolumn{1}{c}{\textbf{SDR (te)}} & \multicolumn{1}{c}{\textbf{eSTOI (te)}} & {\textbf{PESQ (te)}} & \multicolumn{1}{c}{\textbf{SDRI (vl)}} & \multicolumn{1}{c}{\textbf{SDR (vl)}} & \multicolumn{1}{c}{\textbf{eSTOI (vl)}} & \multicolumn{1}{c}{\textbf{PESQ (vl)}} \\ \midrule
\multirow{4}{*}{\makecell[c]{Baseline Generalist,\\ trained from $\mathbb{G}$, \\ tested on $\mathbb{S}^{(i)}$\\ $(i\in\{1,\ldots,20\})$}} & L                                 & 9.84                                   & 10.38                                 &             0.70                             & \textbf{1.68}& 11.76                                 & 11.25                                &0.81 &  \textbf{ 2.29}                                      \\
                                    & M                                 & 9.74                                   & 10.25                                 & 0.69&      \textbf{1.59}                                   & 11.34                                  & 10.83                                & 0.79&   2.19                                      \\
                                    & S                                 & 8.75                                   & 9.29                                  &      0.66                                    &      1.50                                    & 10.45                                  & 9.94                                 &     0.76                                    &        \textbf{2.01}                                 \\
                                    & T                                 & 7.89                                   & 8.43                                  & 0.64&     1.42                                     & 9.70                                  & 9.19                                 &     0.74                                    &      1.89                                   \\
\midrule
PSE Model,     & L                                 & 10.13                                  & 10.65                                 &     0.69                                     &                       1.65                   & \textbf{12.12}                                 & \textbf{11.61}                                &            0.81                             &                     \textbf{2.29}                    \\
finetuned from $\Tilde{\mathbb{S}}^{(i)}_\text{YourTTS}$,& M                               & 9.53                                   & 10.04                                 &               0.68                           &   \textbf{1.59}                                     & 11.77                                 & 11.27                                &                 \textbf{0.80}                        &    \textbf{2.22}                                   \\
$|\Tilde{\mathbb{S}}^{(i)}_\text{YourTTS}|=$\textbf{60} sec.& S                                 & 8.69                                   & 9.20                                  & 0.64&                1.47                          & 10.86                                   & 10.35                                &     \textbf{0.77}                                    &                         \textbf{2.01}                \\
$(i\in\{1,\ldots,20\})$& T                                 & 7.84                                    & 8.35                                  &    0.62                                      &       1.38                                   & \textbf{10.83}                                & \textbf{9.61}                                 &                                 0.74        &       1.85                                  \\
\midrule                                    
PSE Model,& L                                 & 9.98                   &     10.49             &                0.68                          &         1.61                                 &        \textbf{12.12}          &      11.57          &      \textbf{0.82}                                   &                        2.27                 \\
finetuned from $\Tilde{\mathbb{S}}^{(i)}_\text{YourTTS}$& M                                 &        9.39           &      9.90             &        0.67                                  &      1.56                                    &         \textbf{11.87}          &      \textbf{11.32}            &    \textbf{0.80}                                     &                            2.19             \\
$|\Tilde{\mathbb{S}}^{(i)}_\text{YourTTS}|=$\textbf{120} sec.& S                                 &         8.56            &     9.07              &        0.64                                  &        1.45                                  &   \textbf{10.93}                 &         \textbf{10.38}         &      \textbf{0.77}                                   &           1.97                              \\
$(i\in\{1,\ldots,20\})$& T                                 &        7.47             &       7.98          &           0.61                               &   1.34                                       &        10.10          &        9.55          &                   0.74                      &       1.80                                  \\
\midrule
PseudoSE +DP \cite{SivaramanA2022jstsp}, & L & \textbf{10.40} & \textbf{10.91} & \textbf{0.72} & 1.62 & & & & \\
trained via SSL,& M & \textbf{10.19} & \textbf{10.70} & \textbf{0.71} & 1.58 & & & & \\
tested on $\mathbb{S}^{(i)}$& S & \textbf{9.88} & \textbf{10.39} & \textbf{0.70} & \textbf{1.55} &  & & \\
 $(i\in\{1,\ldots,20\})$ & T & \textbf{9.40} & \textbf{9.91} & \textbf{0.68} & \textbf{1.49} & & & \\
\bottomrule                          
\end{tabular}
}
\caption{Experimental results for speaker set 1. Best results for each size and model are indicated in bold.}
\label{tab:exp1}
\end{table*}

\begin{table*}[t]
\centering
\resizebox{.81\textwidth}{!}{%
\begin{tabular}{@{}c|c|rrrr|rrrr@{}}
\toprule
\textbf{Experiment \#2}                 & {\textbf{Size}} & \multicolumn{1}{c}{\textbf{SDRI (te)}} & \multicolumn{1}{c}{\textbf{SDR (te)}} & \multicolumn{1}{c}{\textbf{eSTOI (te)}} & {\textbf{PESQ (te)}} & \multicolumn{1}{c}{\textbf{SDRI (vl)}} & \multicolumn{1}{c}{\textbf{SDR (vl)}} & \multicolumn{1}{c}{\textbf{eSTOI (vl)}} & \multicolumn{1}{c}{\textbf{PESQ (vl)}} \\ \midrule
\multirow{4}{*}{\makecell[c]{Baseline Generalist,\\ trained from $\mathbb{G}$, \\ tested on $\mathbb{S}^{(i)}$\\ $(i\in\{21,\ldots,26\})$}} & L                                 & 11.26                                  & 10.75                                 &   \textbf{0.69}                                      &            1.38                              & 14.23                                 & 13.20                                &               \textbf{0.75}                         &     1.71                                    \\
                                    & M                                 & 10.12                                  & 9.61                                  &   0.67                                       &     1.32                                     & 12.43                                 & 11.40                                &         0.73                                &      1.60                                   \\
                                    & S                                 & 10.31                                  & 9.80                                  & 0.65&     1.28                                     & 12.87                                  & 11.83                                &        \textbf{0.72}                                &     \textbf{1.55}                                    \\
                                    & T                                 & 8.88                                   & 8.38                                  &       0.64                                   &         1.24                                 & 11.29                                 & 10.26                                &                  \textbf{0.71}                       &              1.46                           \\ 
\midrule                                    
PSE Model,& L                                 & \textbf{11.81}                                  & \textbf{11.31}                                 &       \textbf{0.69}                                 &       \textbf{1.43}                                  & \textbf{14.69}                                & 13.66                                &       \textbf{0.75}                                  &                              \textbf{1.75}           \\
finetuned from $\Tilde{\mathbb{S}}^{(i)}_\text{AudioLM}$& M                                 & \textbf{11.36}                                  & \textbf{10.86}                                &\textbf{0.69} & \textbf{1.38}                                          & \textbf{14.08}                                 & 13.05                                &      \textbf{0.75}                                 &                               \textbf{1.69}          \\
$|\Tilde{\mathbb{S}}^{(i)}_\text{AudioLM}|=$\textbf{21} sec.& S                                 & \textbf{10.73}                                  & \textbf{10.22}                                &               \textbf{0.66}                           &        \textbf{1.31}                                  & \textbf{13.37}                                 & 12.34                                &                     \textbf{0.72}                    &          1.57                               \\
$(i\in\{21,\ldots,26\})$& T                                 & \textbf{9.98}                                   & \textbf{9.47}                                 &        \textbf{0.65}                                  &    \textbf{1.26}                                      & \textbf{12.48}                                 & \textbf{11.45}                                &                               \textbf{0.71}        &        \textbf{1.48}                                 \\
\midrule                                    
PSE Model,& L                                 & 10.99                                  & 10.48                                 & 0.64&    1.37                                      & 14.30                                 & \textbf{13.81}                                &                     0.74                    &    1.68                                     \\
finetuned from $\Tilde{\mathbb{S}}^{(i)}_\text{YourTTS}$& M                                 & 10.51                                  & 10.00                                 & 0.67&        1.33                                  & 13.87                                 & \textbf{13.37}                                &     0.73                                    &                 1.63                        \\
$|\Tilde{\mathbb{S}}^{(i)}_\text{YourTTS}|=$\textbf{21} sec.& S                                 & 9.89                                   & 9.38                                  & 0.64&                       1.27                   & 13.03                                 & \textbf{12.54}                                &    0.70                                     &                          1.51               \\
$(i\in\{21,\ldots,26\})$& T                                 & 9.25                                   & 8.75                                  & 0.64&    1.24                                      & 11.88                                 & 11.39                                &         0.70                                &     1.44                                    \\
\midrule
PSE Model,& L                                 & 11.40                                  & 10.89                                 &       \textbf{0.69}                                  &                1.39                          & 12.89                                  & 12.45                                &         0.74                                &                                1.71         \\
finetuned from $\Tilde{\mathbb{S}}^{(i)}_\text{YourTTS}$& M                                 & 10.78                                  & 10.27                                  &0.67 &                  1.35                        & 12.54                                   & 12.10                                &        0.73                                 &                 1.66                        \\
$|\Tilde{\mathbb{S}}^{(i)}_\text{YourTTS}|=$\textbf{60} sec.& S                                 & 10.22                                  & 9.72                                  &         0.63                                 &     1.30                                     & 11.69                                 & 11.25                                  &       0.70                                  &     \textbf{1.55}                                    \\
$(i\in\{21,\ldots,26\})$& T                                 & 9.42                                   & 8.91                                  & 0.62&         1.23                                 & 10.83                                 & 10.39                                &        0.68                                 &      1.40                                   \\
\bottomrule
\end{tabular}
}
\caption{Experimental results for speaker set 2. Best results for each size and model are indicated in bold}
\label{tab:exp2}
\end{table*}

\subsection{Experiment \#1: Comparison with other PSE methods}

The goal of the experiments with $\Tilde{\mathbb{S}}^{(1:20)}$ is to show the benefit of synthetic data augmentation compared to speaker-agnostic models. We generate two sets using YourTTS, with length 60 or 120 seconds, in order to determine the minimal amount of synthesized data needed to achieve enhancement performance comparable to the generalist's. These experiments use the same test set as generalist and specialist models in \cite{SivaramanA2022jstsp}, allowing for a direct comparison. 

\subsection{Experiment \#2: Comparison between NSS schemes}

Experiments with $\Tilde{\mathbb{S}}^{(21:26)}$ aim to determine the impact of the quality of the synthesized speech as well as the amount of high-quality data needed for fine-tuning as opposed to lower-quality counterparts. To this end, we finetune the generalist with the two synthesized sets $\Tilde{\mathbb{S}}^{(21:26)}_\text{YourTTS}$ and $\Tilde{\mathbb{S}}^{(21:26)}_\text{AudioLM}$, resulting in $6\times 2$ PSE models.


\section{Results and Discussion}
\label{sec:results}

\subsection{NSS Systems Performance}

Evaluating the quality of the synthesized utterances $\Tilde{\mathbb{S}}$ is infeasible due to the subjective nature of assessment and the difficulty in conducting a listening test on a large-scale synthesized dataset. Additionally, reliable objective metrics, such as PESQ and STOI, cannot be applied due to the lack of clean references. 

To verify our assumption about the quality difference between the two NSS models, we use non-intrusive objective quality evaluation metrics. Instead of the mean opinion scores (MOS), we use an open-source neural network MOS estimator \cite{NittagG2021nisqa}, with which we can indirectly compare the speech quality of the synthesized results.  
In addition, to compare the personality-preservation performance of the differently synthesized speech, we extract x-vectors with the neural speaker encoder developed by the SpeechBrain project \cite{speechbrain}.
Then, we compute the average cosine similarity between the synthesized utterances' x-vector and that of the ground-truth target speaker.
As can be seen in Table \ref{tab:synth_quality}, the estimated MOS scores of YourTTS samples are lower than the quality of AudioLM samples.
Furthermore, the high cosine similarity indicates that speaker personality is better preserved using AudioLM. 

Note that these results do not necessarily imply the overall performance of the two NSS systems in comparison, as their direct comparison is unfair due to various reasons. For example, the AudioLM's demo examples we collected from the authors' website might not correctly represent the model's overall quality. Meanwhile, the test speakers in $\mathbb{S}^{(1:20)}$ were already seen by YourTTS during its pretraining, so the results might be better than its actual performance on unseen speakers. 
However, if we limit the comparison to examples used in this paper, it is convincing that AudioLM's examples are better than YourTTS's. Next, we will see their influence on PSE. 

\subsection{Experiment \#1}
The results of Experiment \#1 are summarized in Table \ref{tab:exp1}. First, on the test set, we can see that the proposed PSE models, in general, underperform the baseline generalist model except for a few large model cases. Meanwhile, their performance on the validation set is indeed consistently better than the baseline. Considering that the validation set was also built based on YourTTS's synthesized utterances, the noticeable performance improvement on the validation set appears to signify an overfitting case. In other words, the finetuning-based personalization was done on the wrong speaker due to the mismatch between the target speaker and the synthesized speech in terms of personality. This trend does not change even if we double the size of synthesized speech to 120 seconds---the larger augmented set actually worsens the situation. For comparison, we reproduce the results of one of the self-supervised learning (SSL) methods from \cite{SivaramanA2022jstsp}, namely pseudo speech enhancement (PseudoSE) and data purification (DP), which shows the best results even though they do not use any clean speech of the target speaker. Experiment \#1 results suggest that a well-designed SSL can outperform the finetuning-based PSE if the data augmentation does not maintain the personality of the target speaker. 

\subsection{Experiment \#2}
Experiment \#2's results are shown in Table \ref{tab:exp2}. We see that the NSS-based data augmentation, if better NSS results from AudioLM are used, improves the PSE performance in every way with a significant margin (more than 1 dB in terms of SDR). An interesting observation we can make here is that there is a clear gap between the PSE models depending on which augmented set they are finetuned from, i.e., $\Tilde{\mathbb{S}}^{(i)}_\text{AudioLM}$ vs. $\Tilde{\mathbb{S}}^{(i)}_\text{YourTTS}$. This result resonates with Table \ref{tab:synth_quality}, where the AudioLM examples showed better speech quality and personalization performance in all metrics. In this smaller subset with six speakers, it appears that YourTTS adapts to the target speakers better than it does to the first speaker set in Experiment \#1. Yet, there is a clear gap between the two NSS systems' impact on PSE. Note that the direct comparison to the PseudoSE+DP results shown in Table \ref{tab:exp1} is impossible due to the mismatch of the two subsets.

\section{Conclusion}

Our work investigated the potential of neural speech synthesis methods for adapting speech enhancement models towards a particular speaker. We employed two off-the-shelf neural synthesis systems (YourTTS and AudioLM) to synthesize new speech utterances as if they were spoken by the target speaker. Because YourTTS and AudioLM vary in performance depending on the conditioning mechanism and their own model capacity, we assessed the output speech quality and personality-preservation of both systems using non-intrusive metrics. Our experiments demonstrate that speech synthesis quality does correlate with usefulness towards PSE. Using the best-quality synthesis dataset, we show that it is possible to implement efficient PSE systems via a simple finetuning approach. 

\vfill\pagebreak

\bibliographystyle{IEEEbib}
\bibliography{mjkim, anakuzne}

\end{document}